\documentclass[conf]{new-aiaa}
\usepackage[utf8]{inputenc}

\usepackage{graphicx,tikz}
\usepackage{amsmath}
\usepackage[version=4]{mhchem}
\usepackage{siunitx}
\usepackage{longtable,tabularx}
\title{Cognitive Preadaptation for Resilient Adaptive Control}

\author{Deepan Muthirayan \footnote{Post-doctoral Researcher, Electrical Engineering and Computer Sciences, deepan.m@uci.edu.} and Pramod P. Khargonekar \footnote{Professor, Electrical Engineering and Computer Sciences, pramod.khargonekar@uci.edu@uci.edu.}}
\affil{University of California, Irvine, CA.}

\begin{document}

\maketitle

\begin{abstract}
In this paper, we investigate a novel control architecture and algorithm for incorporating preadaption functions. We propose a preadaptation mechanism that can augment any adaptive control scheme and improve its resilience. We also propose a preadaptation learner that learns the preadaption function with experience, which removes the complexity of designing and fine tuning the preadaptation function specific to the system to be controlled. Through simulations of a flight control system we illustrate the effectiveness of the preadaptation mechanism in improving the adaptation. We show that the preadaptation mechanism we propose can reduce the peak of the response by as much as $50\%$. The scenarios we present also show that the preadaptation mechanism is effective across a wide range of scenarios suggesting that the mechanism is reliable. 
\end{abstract}

\section{Introduction}

Adaptability is a key capability of intelligent and autonomous systems. Adaptability enables such systems to learn and optimize for better efficiency, performance, reliability, and resiliency to uncertainties and external changes \cite{tschan2016roadmap}. But there are challenges to the design of adaptive systems. In the classical adaptive control setting a well known trade-off in the design of adaptive control algorithms is the trade-off between rate of adaptation and robustness restricting the scope for increasing the rate of adaptation \cite{nguyen2018verification}. Methods for improving the rate of adaptation while maintaining robustness such as L1 adaptive control \cite{hovakimyan2010L1,hovakimyan20111,jafari2013l1,pereida2017high} and others such as \cite{yucelen2012low,yucelen2014improving} have been proposed earlier. 

Over the last two years, we have been exploring a vision for the future of intelligent and autonomous systems as cognitive cyber-physical systems \cite{ppkesweek, ppkacc20}. The main idea is to endow cyber-physical systems with cognitive capabilities such as memory, attention, learning, problem solving, etc. \cite{richardson1996working, tulving1985many, posner2011cognitive, gershman2017reinforcement}. In this paper, we will explore the idea that the human brain functions as a ``preadaptive organ'' endowing the human the ability to adapt proactively by anticipating changes instead of reactively \cite{fuster2014prefrontal, fuster2017prefrontal}. This concept also aligns well with ideas from memory systems and the notion of ``premembering expereince'' as articulated in \cite{nobre}. This preadaptation is a feature of cognitive capabilities of the human brain and thus fits into the future cognitive cyber-physical systems. 

More specifically, in this paper we investigate  control architecture and algorithm for incorporating preadaption functions. Our goal is to investigate the hypothesis that preadaptation allows the closed loop system to adapt more resiliently without increasing the risks of adaptation. In this paper, we propose a preadaptation mechanism that can augment any adaptive control scheme and improve its resilience. For illustration of the idea, we consider the standard MRAC control framework \cite{whitaker1958design}. 

In Section \ref{sec:problemsetting} we describe the problem setting. In Section \ref{sec:controlarch} we introduce the novel control architecture with the preadaptation module. In Section \ref{sec:cogpread} we introduce the preadapation mechanism. Here we discuss the sub-functions that constitute the preadaptation mechanism. Finally, in Section \ref{sec:disc-simresults} we provide simulation examples to illustrate the preadaptation mechanism.

\subsection{Problem Setting}
\label{sec:problemsetting}

We consider the following class of plants:
\begin{equation} 
\dot{x}(t) = A x(t)+B\left(\theta^Tx(t)+u(t)\right)+B_{1,r}r(t), \ x(0) = x_0, y = x_i,
\label{eq:plant}
\end{equation}
where $x$ is the state vector and is assumed to be measurable, and $x \in \mathbb{R}^n$, $x_i$ denotes the $i$th component of the state, $u \in \mathbb{R}$ is the control input, $\theta \in \mathbb{R}^n$ is an unknown parameter vector that belongs to a known compact convex set $\Omega \subset \mathbb{R}^n$, $A \in \mathbb{R}^{n\times n}, B \in \mathbb{R}^{n\times 1}$, the pair ($A, B$) is controllable, $B_{1,r} \in \mathbb{R}^n$, the matrices $A, B, B_{1,r}$ are known and $r(t) \in \mathbb{R}$ is a bounded reference signal. The objective is to choose $u(t)$ such that all signals in the closed-loop system are uniformly bounded and $x(t)$ tracks the state vector of the desired reference model,
\begin{equation}
\dot{x}_r(t) = A_rx_r(t)+B_{1,r}r(t)+B_{2,r}r(t), \ x_r(0) = x_0.
\label{eq:refmodel}
\end{equation}
both in transient and in steady-state, where $B_{2,r} = k_0B$, $A_r = A - BK, K \in \mathbb{R}^{1\times n}$ is a stabilizing control gain, and $k_0 \in \mathbb{R}$. The main goal is to {\it design a preadaptation mechanism for the adaptive controller such that the response of the closed loop system to track the reference model is resilient}.

\subsection{Contribution}

The main contribution of this work is proposing a {\it novel adaptive control architecture} based on {\it cognitive preadaptation} for {\it resiliency in adaptation} and designing a preadaptation mechanism for the setting in Section \ref{sec:problemsetting}. 

\section{Preadaptation and Adaptive Control Algorithm}
\label{sec:controlarch}

\begin{figure}
\begin{center}
\begin{tikzpicture}[scale = 0.7]

\draw [draw=blue, fill=blue, fill opacity = 0.1, rounded corners, thick] (-3, 1.5) rectangle (-1, 2.5);
\draw (-2, 2) node [align=center] {\tiny Baseline Control};

\draw [draw=blue, fill=blue, fill opacity = 0.1, rounded corners, thick] (-3, -0.5) rectangle (-1, 0.5);
\draw (-2, 0) node [align=center] {\tiny Control Law};

\draw [draw=blue, fill=blue, fill opacity = 0.1, rounded corners, thick] (2, -0.5) rectangle (4, 0.5);
\draw (3, 0) node [align=center] {\tiny Plant};

\draw [draw=blue, fill=blue, fill opacity = 0.1, rounded corners, thick] (5.5, -0.5) rectangle (7.5, 0.5);
\draw (6.5, 0) node [align=center] {\tiny Output};

\draw [draw=blue, fill=blue, fill opacity = 0.1, rounded corners, thick] (-3, -2.5) rectangle (-1, -1.5);
\draw (-2, -2) node [align=center] {\tiny Ref. Model};

\draw [draw=blue, fill=blue, fill opacity = 0.1, rounded corners, thick] (-3, -3.5) rectangle (-1, -4.5);
\draw (-2, -4) node [align=center] {\tiny Adaptation Law};

\draw [draw=blue, fill=blue, fill opacity = 0.1, rounded corners, thick] (5.5, -2.5) rectangle (7.5, -1.5);
\draw (6.5, -2) node [align=center] {\tiny Velocity Estimator};

\draw [draw=blue, fill=blue, fill opacity = 0.1, rounded corners, thick] (5.5, -3.5) rectangle (7.5, -4.5);
\draw (6.5, -4) node [align=center] {\tiny Attention$(e,\hat{\dot{e}})$};

\draw [draw=blue, fill=blue, fill opacity = 0.1, rounded corners, thick] (2, -3.5) rectangle (4, -4.5);
\draw (3, -4) node [align=center] {\tiny Preadaptation};

\draw (0.5,0) node [draw=blue, circle, thick] (a) {$+$};
\draw (0.5,-2) node [draw=blue, circle, thick] (b) {$-$};
\draw (9,-1) node [draw=blue, circle, thick](e) {$-$};

\draw (-4.5, 2) node [anchor = center] (c) {\scriptsize $r$};
\draw (-3.75, -2) node [anchor = center] (k) {\scriptsize $r$};
\draw (4.5, 0) node [anchor = south] {\scriptsize $x$};
\draw (7.75, 0) node [anchor = south] {\scriptsize $y$};
\draw (-0.25, -2) node [anchor = south] {\scriptsize $x_r$};
\draw (11, -1.3) node [anchor = south] (l) {\scriptsize $x_r(i)$};
\draw (-0,-3.75 ) node [anchor = south] {\scriptsize $e_v$};
\draw (-3.5, -4) node [anchor = south] {\scriptsize $\hat{\theta}$};
\draw (-0.5, 2) node [anchor = south] {\scriptsize $u_{bl}$};
\draw (-0.5, 0) node [anchor = south] {\scriptsize $u_{ad}$};
\draw (1.5, 0) node [anchor = south] {\scriptsize $u$};
\draw (8, -1) node [anchor = south] (d) {\scriptsize $e$};
\draw (6.5, -3) node [anchor = east] (h) {\scriptsize $\hat{\dot{e}}$};
\draw (1.5, -4.25) node [anchor = south] {\scriptsize $\hat{\theta}_I$};
\draw (7, -4.75) node [anchor = north] {\scriptsize Preadaptation};
\draw (-2.4, 0.75) node [anchor = south] {\scriptsize Adaptive Control};
\draw (5,-1) node (f) {}; \draw (2.5,-2) node (g) {}; \draw (5,-3) node (i) {};\draw (5,-4) node (j) {};

\draw [->,thick] (4,0) -- (5,0) -- (5,3) -- (-2,3) -- (-2,2.5);
\draw [->,thick] (5,0) -- (5.5,0);
\draw [->,thick] (-1,2) -- (0.5,2) -- (a);
\draw [->,thick] (-1,0) -- (a); \draw [->,thick] (a) -- (2,0);
%\draw [->,thick] (a) -- (0.5,-1) -- (-3.5,-1) -- (-3.5,-2) -- (-3,-2);
\draw[ ->,thick] (k) -- (-3,-2);
\draw[ ->,thick] (l) -- (e);
\draw [->,thick] (4,0) -- (5,0) -- (5,-2) -- (b);
\draw [->,thick] (-1,-2) -- (b); \draw [->,thick] (b) -- (0.5,-3.75) -- (-1,-3.75); 
\draw [->,thick] (-3,-4) -- (-4,-4)  -- (-4,-0.25) -- (-3,-0.25);
\draw [->,thick] (5,-2) -- (5,-5.25) -- (-4.5,-5.25) -- (-4.5,0) -- (-3,0);
\draw [->,thick] (c) -- (-4.5,0.25) -- (-3,0.25);
\draw [->,thick] (c) -- (-3,2); \draw [->,thick] (7.5,0) -- (9,0) -- (e);
%\draw [->,thick] (c) -- (-4.5,3.5) -- (10,3.5) -- (10,-1) -- (e); 
\draw [-> ,thick] (e) -- (f) -- (2.5,-1) -- (g) -- (2.5,-3.5);
\draw [->,thick] (6.5,-1) -- (6.5,-1.5); \draw [->,thick] (6.5,-2.5) -- (6.5,-3.5); \draw [->,thick]  (8,-1) -- (8,-4) -- (7.5,-4);
 \draw [->,thick] (h) -- (i) -- (3.5,-3) -- (3.5,-3.5); \draw [->,thick] (2,-4.25) -- (-1,-4.25); 
\draw [->,thick] (-4,-4) -- (-4,-5) -- (3,-5) -- (3,-4.5); \draw [->,thick] (5.5,-4) -- (j) -- (4,-4);

\draw [- ,draw = orange, fill = orange, fill opacity = 0.05, dashed] (1.75,-0.75) -- (8.25,-0.75) -- (8.25,-4.75) -- (1.75,-4.75) -- cycle; 

\draw [-,draw = green, fill = green, fill opacity = 0.05, dashed] (-0.5, 0.75) -- (-4.25,0.75) -- (-4.25,-4.75) -- (-0.5,-4.75) -- cycle; 

\end{tikzpicture}
\caption{Control Architecture: Adaptive control with Preadaptation}
\label{fig:contarch}
\end{center}
\end{figure}
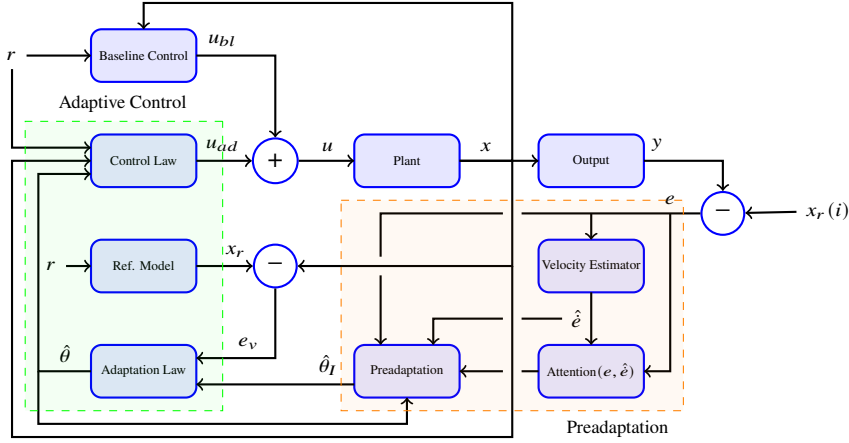

The proposed adaptive control architecture with preadaptation is shown in Fig. \ref{fig:contarch}. The adaptive control module is the standard adaptive control module and is discussed below. The preadaptation mechanism has an attention function that can identify the occurence of a sudden disturbance in $\theta$ by observing the deviation of the error in the response. When such an occurence is identified by the attention function, the preadaptation function reinitializes the adaptation mechanism that outputs the estimate $\hat{\theta}$ of the unknown parameter $\theta$, with $\hat{\theta}_I$. Part of the preadaptation mechanism is a {\it preadaptation learner} that learns a suitable preadaptation function with experience.

\subsection{Control Law and Adaptation Algorithm}
\label{sec:contlaw}

The final control input $u$ is the summation of the baseline control $u_{bl}$ and the adaptive control input $u_{ad}$:
\begin{equation}
u = u_{bl} + u_{ad}.
\label{eq:controlinput}
\end{equation} 
The {\it baseline control} for the system in Eq. \eqref{eq:plant} is given by %has to track the reference model in Eq. \eqref{eq:refmodel} 
\begin{equation}
u_{bl} = -Kx
\label{eq:baseline}
\end{equation}
The {\it adaptive control law} and the {\it adaptation law} are the standard laws for the setting described in Section \ref{sec:problemsetting} (please see \cite{jafari2013l1}), given by,
\begin{equation}
u_{ad} = -\hat{\theta}^Tx +k_0r(t),
\label{eq:adaptivecontrol}
\end{equation}
where the notation $v^T$ denotes the transpose of the vector $v$ and
\begin{equation}
\dot{\hat{\theta}} = \gamma xe_v^TPB, e_v = x - x_r,
\label{eq:adaptivecontrol}
\end{equation}
where $P > 0$ and is the solution of the Lyapunov equation $PA_r + PA^T_r = -I$.
 
\subsection{Preadaptation Mechanism}
\label{sec:cogpread}

The preadaptation mechanism we propose has two sub-functions: (i) an {\it attention function}, and (ii) a {\it preadaptation function}. Below, we discuss the two functions in detail.

\subsubsection{Attention Function}

Denote the output of the velocity estimator in Fig. \ref{fig:contarch} by $\hat{\dot{e}}$, where $\hat{\dot{e}}$ is an estimate of $\dot{e}$ and $e$ is the output error. The attention function flags the occurence of a disturbance in the unknown parameter $\theta$ at the moment $(|e| - c_e)$ crosses zero from below and if $\vert\hat{\dot{e}}\vert > c_{ed} > 0$ and similarly flags that the system has recovered after a disturbance exactly when the opposite happens. Denote the indicator of the instance when $(|e| - c_e)$ crosses zero from below ($(|e| - c_e) \uparrow 0$) and $\vert\hat{\dot{e}}\vert > c_{ed} > 0$ by $E_u$. Then
\begin{equation}
E_u = \left\{ \begin{array}{cc} 1 & \text{when} \ (|e| - c_e) \uparrow 0 \ \text{and} \ \vert\hat{\dot{e}}\vert > c_{ed} \\ 0 & \text{otherwise}. \end{array} \right.  \nonumber 
\end{equation}
Similarly, denote the indicator of the instance when $(|e| - c_e)$ crosses zero from above ($(|e| - c_e) \downarrow 0$) and $\vert \hat{\dot{e}} \vert < c_{ed}$ by $E_d$. Then
\begin{equation}
E_d = \left\{ \begin{array}{cc} 1 & \text{when} \ (|e| - c_e) \downarrow 0 \ \text{and} \ \vert\hat{\dot{e}}\vert < c_{ed} \\ 0 & \text{otherwise}. \end{array} \right.  \nonumber 
\end{equation}
Thus the attention function $\text{Att}(e,\hat{\dot{e}})$ is given by
\begin{equation}
\text{Att}(e,\hat{\dot{e}}) = \left\{ \begin{array}{cc} 1, & \ \text{if} \ E_u (\text{OR}) E_d = 1 \\ 0 & \text{otherwise}. \end{array}\right.
\label{eq:attention}
\end{equation}
where $c_e > 0$ and $c_{ed} > 0$ are constants. Essentially, the attention function indicates the occurence of a disturbance when the magnitude of the error in the output crosses a threshold and at a rate that exceeds a certain threshold. And similarly indicates that the closed loop system is nearly tracking the reference model when the magnitude of the error drops below a threshold and at a rate less than a certain threshold. The mechanism we propose is also easy to implement, because the only parameters that would have to be fine tuned are the $c_e$ and $c_{ed}$ parameters and the simulations reveal that it is effective. 

\subsubsection{Preadaptation Function}

The preadaptation function computes $\hat{\theta}_I$ to reinitialize $\hat{\theta}$ to $\hat{\theta}_I$ whenever the attention functions identifies the occurence of a disturbance, i.e., when $\text{Att} = 1$ and $E_u = 1$. In this work we choose the  function to compute $\hat{\theta}_I$ to be a two layer neural network given by the weights $W_{pa}$ and $V_{pa}$, where $W_{pa}$ and $V_{pa}$ are matrices of appropriate dimensions. Thus, the output $\hat{\theta}_I$ is computed by
\begin{equation}
\hat{\theta}_I = W^T_{pa}\sigma\left(V^T_{pa}\left[e \ \ \vert\hat{\dot{e}}\vert\right]^T\right), \nonumber 
\end{equation}
and the final action of the preadaptation function is given by
\begin{equation}
\left\{ \begin{array}{cc} \hat{\theta} \leftarrow \hat{\theta}_I & \ \text{When} \ \text{Att} = 1 \ \text{and} \ E_u = 1, \\  \text{No action} & \text{otherwise}, \end{array} \right.  \nonumber 
\end{equation}
where $\hat{\theta} \leftarrow \hat{\theta}_I$ denotes the action of $\hat{\theta}$ being reinitialized to $\hat{\theta}_I$. We choose a general function such as a neural network (NN) for the preadaptation function because this allows the preadaptation function to be learned with experience. This reduces the complexity of designing and fine tuning the preadaptation function specific to the system to be controlled.

\subsubsection{Learning the Preadaptation function}

A suitable function to compute the optimal $\hat{\theta}_I$ to reinitialize $\hat{\theta}$ can be learnt by fine tuning the weights $W_{pa}$ and $V_{pa}$ based on how effective the reinitializing was. Hence, the performance metric we choose for updating the preadaptation function should reflect how resilient the adaptation was after reinitializing $\hat{\theta}$ to $\hat{\theta}_I$. In this work we choose the performance metric to be the integral of the magnitude of error $e$ from the instance when the attention mechanism identifies an event of disturbance to the instant when system is adjudged to be tracking the reference model, i.e., when $E_u = 1$ and $E_d = 1$ respectively. Denote the respective time instants by $t_u$ and $t_d$. Then the performance metric for the preadaptation function is given by
\begin{equation}
E = \int_{t_u}^{t_d} \vert e \vert d\tau.
\label{eq:perf-pap}
\end{equation}

The preadaptation function is updated after every adpatation phase as demarked by $E_u = 1$ and $E_d = 1$. When the attention mechanism flages $E_d = 1$, notifying that the system has recovered and is tracking the reference model, the performance of preadaptation is computed as in Eq. \ref{eq:perf-pap} and the weights $W_{pa}$ and $V_{pa}$ are fine tuned by the gradient of $E$ as given below:
\begin{align}
& W_{pa}\vert_{t_d} \leftarrow W_{pa}\vert_{t_u} - \gamma_{pa} \left(\frac{\partial E}{\partial W_{pa}}\right)^T, \nonumber \\
&  V_{pa}\vert_{t_d} \leftarrow V_{pa}\vert_{t_u} - \gamma_{pa} \left(\frac{\partial E}{\partial V_{pa}}\right)^T.
\label{eq:learn-alg-pap} 
\end{align}
The gradient based update allows the preadapation function to be learned with experience that is effective for the specific system to be controlled.

\subsubsection{Gradient Calculation}

In this section we derive $\left(\frac{\partial E}{\partial W_{pa}}\right)^T$ and $\left(\frac{\partial E}{\partial V_{pa}}\right)^T$. Consider the following dynamics:
\begin{equation}
\dot{\hat{\theta}} = g(\hat{\theta},e_v,x_r), \dot{e}_v = h(\hat{\theta},e_v,x_r), \dot{x}_r = f(\hat{\theta},e_v,x_r). \nonumber 
\end{equation}
We will specify these functions later. Let the values of $\hat{\theta}$ and $e_v$ at the instant $t_u$ be denoted by $\hat{\theta}_0$ and $e_{v,0}$ respectively. From this definition it follows that
\begin{equation}
e_v(t) = e_{v,0} + \int_{t_u}^{t} h(\hat{\theta}, e_v, x_r)d\tau, \ \hat{\theta}(t) = \hat{\theta}_{0} + \int_{t_u}^{t}  g(\hat{\theta}, e_v, x_r) d\tau. \nonumber 
\end{equation}
In the deriviation that follows it is understood that the partial derivatives are of appropriate dimensions. Using the fact that $x_r$ is independent of $\hat{\theta}_0$ and taking the partial deriative w.r.t $\hat{\theta}_0$ we get that
\begin{equation}
\frac{\partial e_v(t)}{\partial \hat{\theta}_0} = \int_{t_u}^{t} \left(\frac{\partial h}{\partial \hat{\theta}}\frac{\partial \hat{\theta}}{\partial \hat{\theta}_0} + \frac{\partial h}{\partial e_v}\frac{\partial e_v}{\partial \hat{\theta}_0} \right) d\tau, \ \frac{\partial \hat{\theta}(t)}{\partial \hat{\theta}_0} = I+ \int_{t_u}^{t} \left(\frac{\partial g}{\partial \hat{\theta}}\frac{\partial \hat{\theta}}{\partial \hat{\theta}_0} + \frac{\partial g}{\partial e_v}\frac{\partial e_v}{\partial \hat{\theta}_0} \right) d\tau. \nonumber 
\end{equation}
We note that $\hat{\theta}_0 = \hat{\theta}_I$. Hence
\begin{equation}
\frac{\partial e_v(t)}{\partial \hat{\theta}_I} = \int_{t_u}^{t} \left(\frac{\partial h}{\partial \hat{\theta}}\frac{\partial \hat{\theta}}{\partial \hat{\theta}_I} + \frac{\partial h}{\partial e_v}\frac{\partial e_v}{\partial \hat{\theta}_I} \right) d\tau, \ \frac{\partial \hat{\theta}(t)}{\partial \hat{\theta}_I} = I+ \int_{t_u}^{t} \left(\frac{\partial g}{\partial \hat{\theta}}\frac{\partial \hat{\theta}}{\partial \hat{\theta}_I} + \frac{\partial g}{\partial e_v}\frac{\partial e_v}{\partial \hat{\theta}_I} \right) d\tau. \nonumber 
\end{equation}
Differentiating w.r.t $t$ we get that 
\begin{equation}
\frac{\partial \dot{e}_v}{\partial \hat{\theta}_I} =  \frac{\partial h}{\partial \hat{\theta}}\frac{\partial \hat{\theta}}{\partial \hat{\theta}_I} + \frac{\partial h}{\partial e_v}\frac{\partial e_v}{\partial \hat{\theta}_I}, \ \frac{\partial \dot{\hat{\theta}}}{\partial \hat{\theta}_I} = \frac{\partial g}{\partial \hat{\theta}}\frac{\partial \hat{\theta}}{\partial \hat{\theta}_I} + \frac{\partial g}{\partial e_v}\frac{\partial e_v}{\partial \hat{\theta}_I}. \nonumber 
\end{equation}
Combining both equations in to a single equation, they can be written as
\begin{equation}
\left[ \begin{array}{c} \frac{\partial \dot{e}_v}{\partial \hat{\theta}_I} \\ \\ \frac{\partial \dot{\hat{\theta}}}{\partial \hat{\theta}_I} \end{array} \right] = \left[ \begin{array}{cc}  \frac{\partial h}{\partial e_v} &  \frac{\partial h}{\partial \hat{\theta}} \\ & \\ \frac{\partial g}{\partial e_v} & \frac{\partial g}{\partial \hat{\theta}} \end{array} \right] \left[ \begin{array}{c} \frac{\partial e_v}{\partial \hat{\theta}_I} \\ \\ \frac{\partial \hat{\theta}}{\partial \hat{\theta}_I} \end{array} \right].
\label{eq:dot-ev-hth-der-genexp}
\end{equation}
For the setting described in Section \ref{sec:problemsetting} we have that
\begin{equation}
h(\hat{\theta}, e_v, x_r) = A_re_v +B\left(-\hat{\theta}^T(e_v + x_r) + \theta^T(e_v + x_r) \right),  g(\hat{\theta}, e_v, x_r) = \gamma (e_v + x_r) e_v^TPB. \nonumber 
\end{equation}
From here it follows that
\begin{equation}
\frac{\partial h}{\partial e_v}  = A_r + B(\theta - \hat{\theta})^T, \ \frac{\partial h}{\partial \hat{\theta}} = -B(e_v+x_r)^T, \ \frac{\partial g}{\partial e_v}  = \gamma e^T_vPBI + \gamma (e_v + x_r)B^TP^T, \ \frac{\partial g}{\partial \hat{\theta}} = 0.  \nonumber 
\end{equation}
Substituting the above two expressions in Eq. \eqref{eq:dot-ev-hth-der-genexp} we get that
\begin{equation}
\left[ \begin{array}{c} \frac{\partial \dot{e}_v}{\partial \hat{\theta}_I} \\ \\ \frac{\partial \dot{\hat{\theta}}}{\partial \hat{\theta}_I} \end{array} \right] = \left[ \begin{array}{cc} A_r + B(\theta - \hat{\theta})^T & -B(e_v+x_r)^T \\ & \\  \gamma e^T_vPBI + \gamma (e_v + x_r)B^TP^T & 0 \end{array} \right] \left[ \begin{array}{c} \frac{\partial e_v}{\partial \hat{\theta}_I} \\ \\ \frac{\partial \hat{\theta}}{\partial \hat{\theta}_I} \end{array} \right].
\label{eq:dot-ev-hth-der-exp}
\end{equation}
Let 
\begin{equation}
\Pi(e_v,x_r,\theta-\hat{\theta}) = \left[ \begin{array}{cc} A_r + B(\theta - \hat{\theta})^T & -B(e_v+x_r)^T \\ & \\  \gamma e^T_vPBI + \gamma (e_v + x_r)B^TP^T & 0 \end{array} \right]. \nonumber 
\end{equation}
Then
\begin{equation}
\left[ \begin{array}{c} \frac{\partial e_v(t)}{\partial \hat{\theta}_I} \\ \\ \frac{\partial \hat{\theta}(t)}{\partial \hat{\theta}_I} \end{array} \right] = \exp\left\{ \int_{t_u}^t \Pi(e_v,x_r,\theta-\hat{\theta}) \right\} \left[ \begin{array}{c} \frac{\partial e_v(0)}{\partial \hat{\theta}_I} \\ \\ \frac{\partial \hat{\theta}(0)}{\partial \hat{\theta}_I} \end{array} \right].
\label{eq:ev-hth-der-exp}
\end{equation}
In our case, $e$ is the $i$th component of $e_v$. Hence,
\begin{equation}
\frac{\partial E}{\partial \hat{\theta}_I} = \int_{t_u}^{t_d} \frac{e}{\vert e \vert} \frac{\partial e_v(i)}{\partial \hat{\theta}_I} d\tau, 
\label{eq:E-der-exp}
\end{equation}
where $\frac{\partial e_v(i)}{\partial \hat{\theta}_I}$ is given by Eq. \eqref{eq:ev-hth-der-exp}. In the simulations we implement the integrals in Eq. \eqref{eq:E-der-exp} and Eq. \eqref{eq:ev-hth-der-exp} by an approximate summation. 

Denote the $i$th element of $\hat{\theta}_I$ by $\hat{\theta}_I(i)$. Similarly, denote the element at the $j$th row and $i$ th column of $W_{pa}$ by $W_{pa}(j,i)$ and the $i$th column of $W_{pa}$ by $W_{pa}(:,i)$. The partial derivative
\begin{equation}
\frac{\partial E}{\partial W_{pa}(:,i)} = \frac{\partial E}{\partial \hat{\theta}_I(i)}\frac{\partial \hat{\theta}_I(i)}{\partial W_{pa}(:,i)}. \nonumber
\end{equation}
The term $\frac{\partial E}{\partial \hat{\theta}_I(i)}$ is the $i$th component of $\frac{\partial E}{\partial \hat{\theta}_I}$ and 
\begin{equation}
\frac{\partial \hat{\theta}_I(i)}{\partial W_{pa}(:,i)} = \sigma\left(V^T_{pa}\left[e \ \ \vert\hat{\dot{e}}\vert\right]^T\right)^T.\nonumber
\end{equation}
Substituting these expressions we get that
\begin{equation}
\frac{\partial E}{\partial W_{pa}(:,i)} = \frac{\partial E}{\partial \hat{\theta}_I(i)}\sigma\left(V^T_{pa}\left[e \ \ \vert\hat{\dot{e}}\vert\right]^T\right)^T. \nonumber
\end{equation}
Hence, it follows that
\begin{equation}
\frac{\partial E}{\partial W_{pa}} =  \frac{\partial E}{\partial \hat{\theta}_I}^T\sigma\left(V^T_{pa}\left[e \ \ \vert\hat{\dot{e}}\vert\right]^T\right)^T.
\label{eq:der-E-Wpa}
\end{equation}
The partial derivative
\begin{equation}
\frac{\partial E}{\partial V_{pa}(:,j)} = \sum_i \frac{\partial E}{\partial \hat{\theta}_I(i)}\frac{\partial \hat{\theta}_I(i)}{\partial V_{pa}(:,j)}. \nonumber 
\end{equation}
And
\begin{equation}
\frac{\partial \hat{\theta}_I(i)}{\partial V_{pa}(:,j)} = \frac{\partial W^T_{pa}(:,i)\sigma\left(V^T_{pa}\left[e \ \ \vert\hat{\dot{e}}\vert\right]^T\right)}{\partial V_{pa}(:,j)} = W^T_{pa}(:,i)\frac{\partial \sigma\left(V^T_{pa}\left[e \ \ \vert\hat{\dot{e}}\vert\right]^T\right)}{\partial V_{pa}(:,j)}. \nonumber 
\end{equation}
Note that $\frac{d\sigma(x)}{dx} = \text{diag}\{\sigma(x)\odot(1-\sigma(x))\}$, where $\text{diag}\{v\}$ refers to the matrix with the diagonal entries given by the vector $v$ and the rest of the elements zero, and the notation $\odot$ refers to the element wise product. Denote $\frac{d\sigma(x)}{dx}$ by $\sigma'(x)$. Then
\begin{equation}
\frac{\partial \sigma\left(V^T_{pa}\left[e \ \ \vert\hat{\dot{e}}\vert\right]^T\right)}{\partial V_{pa}(:,j)} = \sigma'\left(V^T_{pa}\left[e \ \ \vert\hat{\dot{e}}\vert\right]^T\right) \frac{\partial \left(V^T_{pa}\left[e \ \ \vert\hat{\dot{e}}\vert\right]^T\right)}{\partial V_{pa}(:,j)}. \nonumber
\end{equation}
That is
\begin{equation}
\frac{\partial \sigma\left(V^T_{pa}\left[e \ \ \vert\hat{\dot{e}}\vert\right]^T\right)}{\partial V_{pa}(:,j)} = \sigma'\left(V^T_{pa}\left[e \ \ \vert\hat{\dot{e}}\vert\right]^T\right)\left[ \underset{j-1 \ \text{columns}}{\underbrace{\mathbf{0}, ...}}, \left[e \ \ \vert\hat{\dot{e}}\vert\right]^T, \underset{n-j \ \text{columns}}{\underbrace{\mathbf{0}, ...}}\right]^T. \nonumber
\end{equation}
For convenience, let us denote the $j$th diagonal element of $\sigma'\left(V^T_{pa}\left[e \ \ \vert\hat{\dot{e}}\vert\right]^T\right)$ by $\sigma'(j)$. Then 
\begin{equation}
\frac{\partial \sigma\left(V^T_{pa}\left[e \ \ \vert\hat{\dot{e}}\vert\right]^T\right)}{\partial V_{pa}(:,j)} = \left[ \underset{j-1 \ \text{columns}}{\underbrace{\mathbf{0}, ...}}, \sigma'(j)\left[e \ \ \vert\hat{\dot{e}}\vert\right]^T, \underset{n-j \ \text{columns}}{\underbrace{\mathbf{0}, ...}}\right]^T. \nonumber 
\end{equation}
Hence
\begin{equation}
\frac{\partial E}{\partial V_{pa}(:,j)} = \sum_i \frac{\partial E}{\partial \hat{\theta}_I(i)}W^T_{pa}(:,i)\left[ \underset{j-1 \ \text{columns}}{\underbrace{\mathbf{0}, ...}}, \sigma'(j)\left[e \ \ \vert\hat{\dot{e}}\vert\right]^T, \underset{n-j \ \text{columns}}{\underbrace{\mathbf{0}, ...}}\right]^T. \nonumber 
\end{equation}
That is 
\begin{equation}
\frac{\partial E}{\partial V_{pa}(:,j)} =  \sum_i \frac{\partial E}{\partial \hat{\theta}_I(i)}W_{pa}(j,i) \sigma'(j)\left[e \ \ \vert\hat{\dot{e}}\vert\right] = \frac{\partial E}{\partial \hat{\theta}_I}W^T_{pa}(j,:)\sigma'(j)\left[e \ \ \vert\hat{\dot{e}}\vert\right] = \sigma'(j)W_{pa}(j,:) \frac{\partial E}{\partial \hat{\theta}_I}^T\left[e \ \ \vert\hat{\dot{e}}\vert\right]. \nonumber 
\end{equation}
Hence, it follows that
\begin{equation}
\frac{\partial E}{\partial V_{pa}} = \sigma'W_{pa}\frac{\partial E}{\partial \hat{\theta}_I}^T\left[e \ \ \vert\hat{\dot{e}}\vert\right].
\label{eq:der-E-Vpa}
\end{equation}

This complets the derivation of the gradient. We note that $\frac{\partial e_v(i)}{\partial \hat{\theta}_I}$ is not calculable exactly because the matrix $\Pi(.)$ is a function of $\theta$ which is an unknown. Hence, we make the approximation where we use
\begin{equation}
\hat{\Pi}(e_v,x_r) = \Pi(e_v,x_r,0) = \left[ \begin{array}{cc} A_r & -B(e_v+x_r)^T \\ & \\  \gamma e^T_vPBI + \gamma (e_v + x_r)B^TP^T & 0 \end{array} \right]. \nonumber 
\end{equation}
in place of $\Pi(.)$ in Eq. \eqref{eq:ev-hth-der-exp}. This approximation introduces an error in the computation of the gradient. We discuss the effect of this approximation in the discussion section.

\section{Simulation Results and Discussion}
\label{sec:disc-simresults}

In this section we discuss preliminary results for a flight control problem.  We consider the control of the flight's longitudinal dynamics. Denote the flight's angle of attack by $\alpha$, the flight's pitch by $q$ and the elevator control input by $u$. The flight's angle of attack and the pitch constitute the state of the system. The output $y$ of the system is its angle of attack, $\alpha$. In addition, we append an integrator, where the output of the integrator is the integral of the error between the output, i.e., the angle of attack and the command signal $r$ that the angle of attack has to track. Denote the output of the integrator by $e_I$, where $e_I = \int{\alpha - r}$. The system equations for the longitudinal dynamics appended with the output of the integrator is
\[ \left[ \begin{array}{c} \dot{e}_I  \\ \dot{\alpha} \\ \dot{q} \end{array} \right] =  \left( \begin{array}{ccc} 0 & 1 & 0  \\ 0 & \frac{Z_\alpha}{mU} & 1 + \frac{Z_q}{mU}  \\ 0 & \frac{M_\alpha}{I_y} & \frac{M_q}{I_y} \end{array} \right)\left[ \begin{array}{c} e_I  \\ \alpha \\ q \end{array} \right] + \left( \begin{array} {c} 0 \\ \frac{Z_\delta}{mU} \\ \frac{M_\delta}{I_y} \end{array} \right)(\theta^Tx + u) + \left( \begin{array} {c} -1 \\ 0 \\ 0 \end{array} \right)r.  \nonumber \]

The system parameters are that of B-$747$ flight. We assume that the flight is traveling at a speed of $U = 274 \ \text{m/s}$ ($0.8$ Mach) and at an altitude of $h = 6000 \ \text{m}$. The flight's mass is $m = 288773 \ \text{Kg}$, and its moment of inertia $I_y = 44877574 \ \text{Kg}\text{m}^2$. The baseline control is the LQR controller. The matrices that define the cost of the LQR controller are given by $Q = I$ and $R = 1$. The values for the other parameters in the system equation above are as follows,
\begin{align}
& \frac{Z_\alpha}{mU} = -0.32, 1 + \frac{Z_q}{mU} = 0.86, \frac{M_\alpha}{I_y} = -0.93, \nonumber \\
& \frac{M_q}{I_y}  = -0.43, \frac{Z_\delta}{mU} = -0.02,  \frac{M_\delta}{I_y} = -1.16.  \nonumber 
\end{align}
The adaptive controller and the preadaptation mechanism constants are the following: $r = 0.1, k_0 = 0, \gamma = 10, c_e = 0.005, c_{ed} = 0.02, \gamma_{pa} = \gamma$ and the number of hidden layer neurons of the neural network that computes $\hat{\theta}_I$ is set as $3$. We use the approximated gradient discussed in the previous section in the preadaptation function update.

\begin{figure}
\centering
\begin{tabular}{ll}
\includegraphics[scale = 0.35]{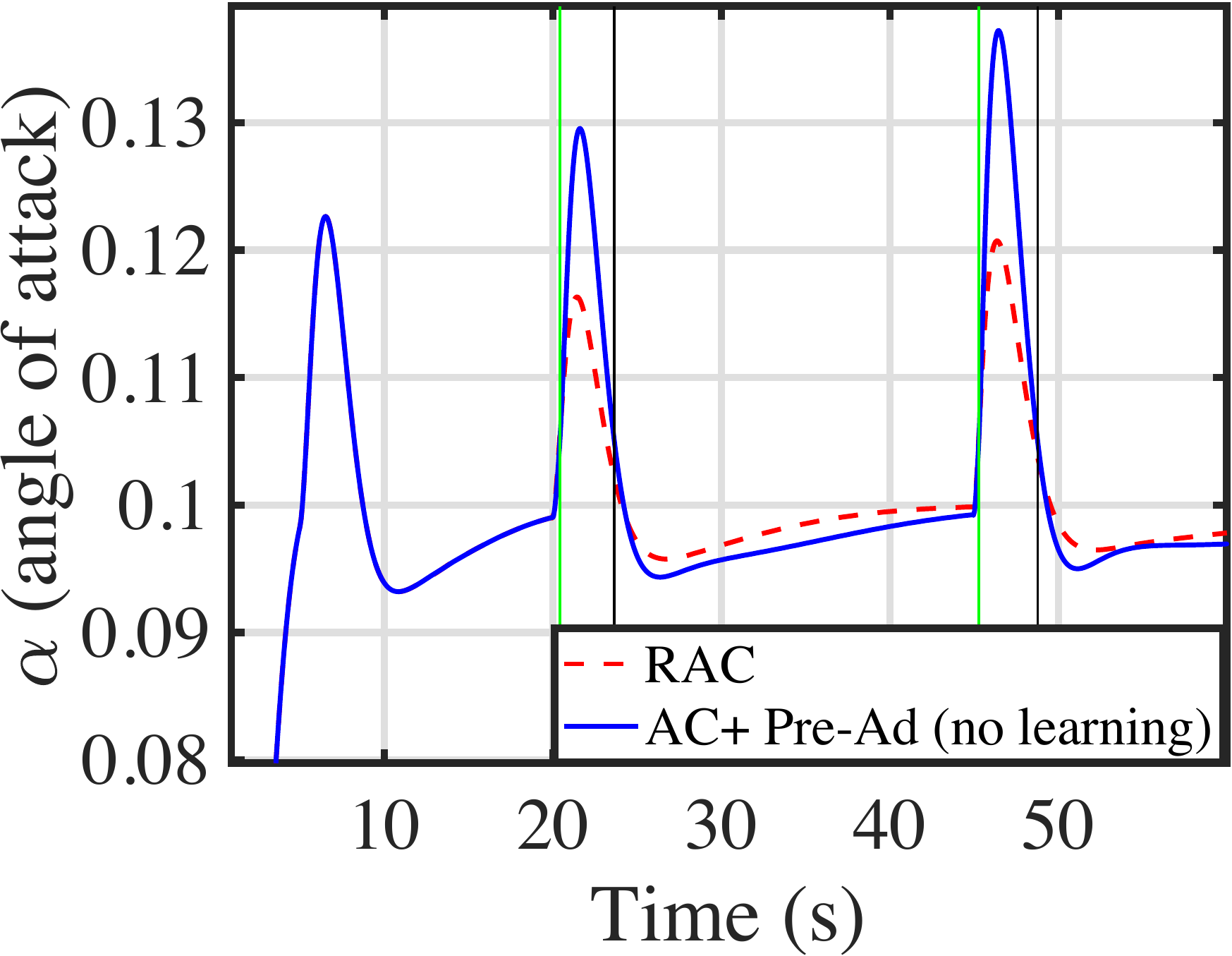} & \includegraphics[scale = 0.35]{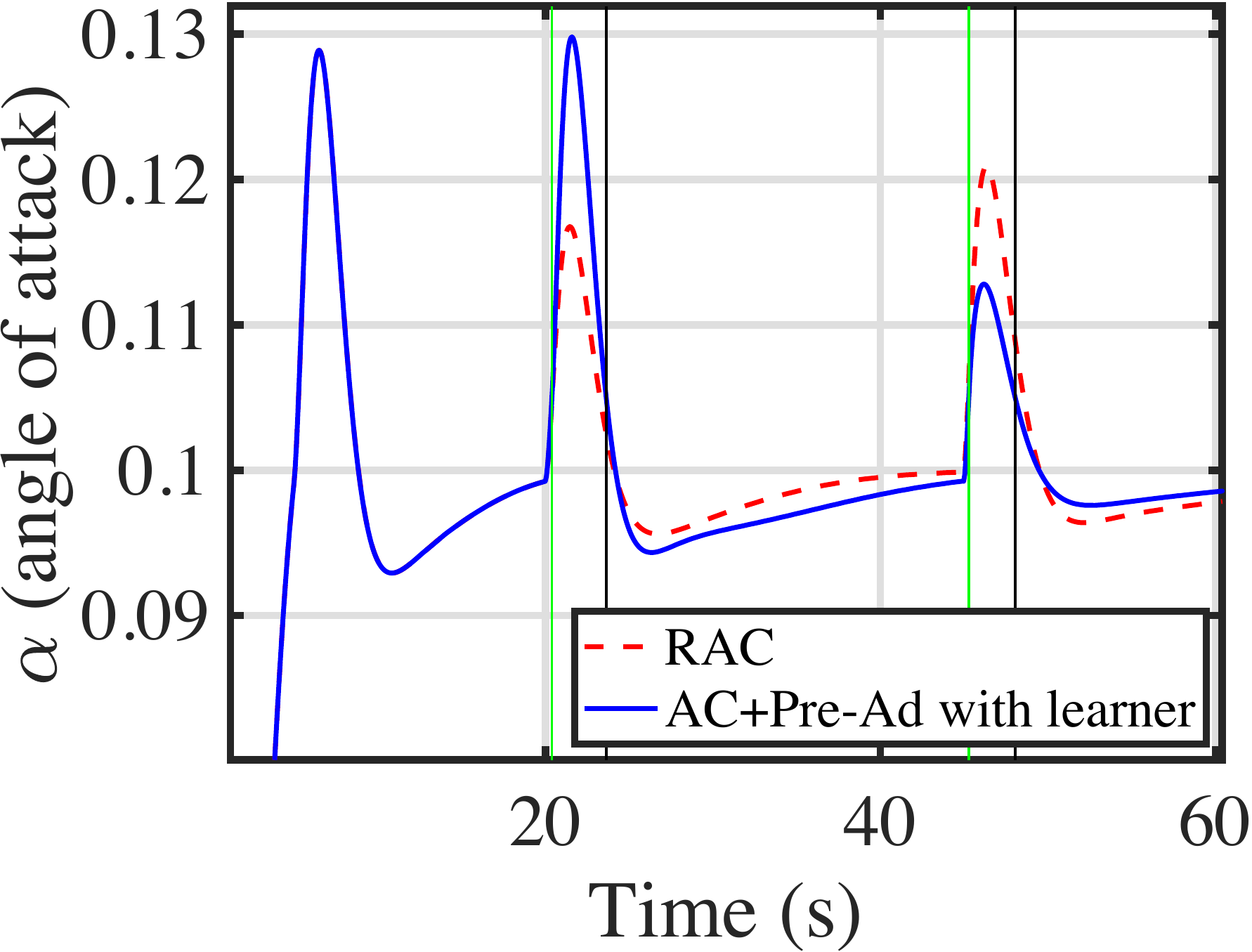} \\
\end{tabular}
\caption{Response of $\alpha$ (angle of attack) for the scenario in Eq. \eqref{eq:theta-scenario-1}. Left: reponse for the adaptive controller with the preadaptation mechanism but without preadaptation learning, right: reponse for the adaptive controller with the preadaptation mechanism and with preadaptation learning. RAC: regular adaptive control, AC: adaptive control. Green line: $E_u = 1$, black line: $E_d = 1$.}
\label{fig:scenario-response-1}
\end{figure}

We present a couple of scenarios to illustrate. In the first scenario we present the unknown parameter changes as follows:
\begin{equation}
\theta = 0.1\mathbf{1}, t \leq 5,  \ \ \theta = 1 \mathbf{1}, 5 < t \leq 20, \ \ \theta = 2\mathbf{1}, 20 < t \leq 45,  \ \ \theta = 4\mathbf{1}, 45 < t \leq 60, 
\label{eq:theta-scenario-1} 
\end{equation}
where $\mathbf{1}$ denotes a vector with all entries as $1$. The response of $\alpha$ for the regular adaptive control without any preadaptation and for the adaptive control with preadaptation are shown in Fig. \ref{fig:scenario-response-1}. The left plot in Fig. \ref{fig:scenario-response-1} shows the response of $\alpha$ for both the controllers when the preadaptation function is randomly initialized and is not fine tuned with experience. The right plot in Fig. \ref{fig:scenario-response-1} shows the response of $\alpha$ for both the controllers when the preadaptation mechanism is randomly initialized and is fine tuned by the learning algorithm described earlier. In the plots, the vertical green line represents the instances when $\text{Att} = 1$ and $E_u = 1$ and the vertical black line represents the instances when $\text{Att} = 1$ and $E_d = 1$. It is evident from the plots that the attention function is correctly able to identify the onset of a disturbance (in this case a shift) in $\theta$ and the instance after which the closed loop system nearly tracks the reference signal $r$ after the onset of a disturbance. 

From the left plot of Fig. \ref{fig:scenario-response-1} it is clear that for the adaptive controller with the preadaptation mechanism but no fine tuning, the adaptation does not improve from one instance to the next instance of disturbance because the preadaptation function is not fine tuned. Whereas for the adaptive controler with the fine-tuning option for the preadaptation function, the improvement in adaptation from one instance to the next instance of disturbance is evident, in this case the disturbances at $t = 20$ and  $t = 45$ respectively, as shown in the right plot of Fig. \ref{fig:scenario-response-1}. The recovery after the disturbance at $t = 45$ with preadaptation fine tuning is much improved and much better than the regular adaptive control with a reduction in peak error by nearly as much as $50 \%$ from the peak error for the regular adaptive controller. This example clearly illustrates the effectiveness of the proposed preadaptation mechanism in improving the recovery following a disturbance. 

In the next scenario we present, the unknown parameter undergoes the following changes: 
\begin{align}
& \theta = 0.1\mathbf{1}, t \leq 5,  \ \ \theta = 1 \mathbf{1}, 5 < t \leq 20, \ \ \theta = 2\mathbf{1}, 20 < t \leq 45,  \ \ \theta = 1\mathbf{1}, 45 < t \leq 70, \nonumber \\
& \theta = 0.1\mathbf{1}, 70 < t \leq 95, \ \theta = 2\mathbf{1}, 95 < t \leq 120, \ \theta = 4\mathbf{1}, 120 < t \leq 140.
\label{eq:theta-scenario-2} 
\end{align}
We first highlight the differences between this scenario and the previous scenario. In this scenario, unlike the previous scenario, the coefficient of the parameter drops after reaching $2$ at the instant $t = 45$, drops further by a factor 10 at $t = 70$ and raises again at $t = 95$. Since the learning algorithm wouldn't have encountered a drop in magnitude of the unknown parameter before $t = 45$, the preadaptation mechanism may not induce a more resilient response after the disturbance at $t = 45$. But we can expect the response to be improved at the next instant when the magnitude drops because it is likely to have learned how to preadapt to such an occurrence by then. 

\begin{figure}
\centering
\begin{tabular}{ll}
\includegraphics[scale = 0.35]{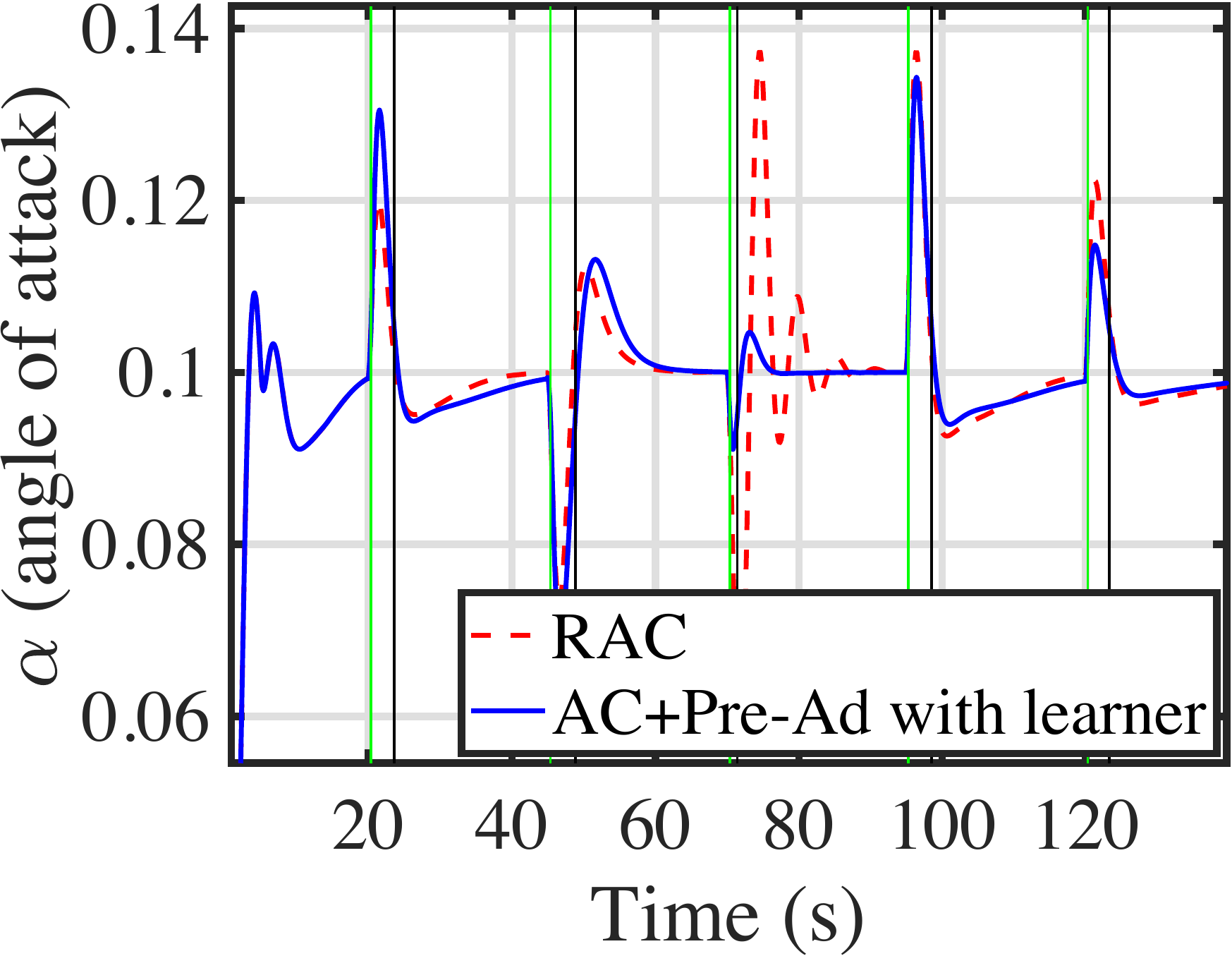} & \includegraphics[scale = 0.35]{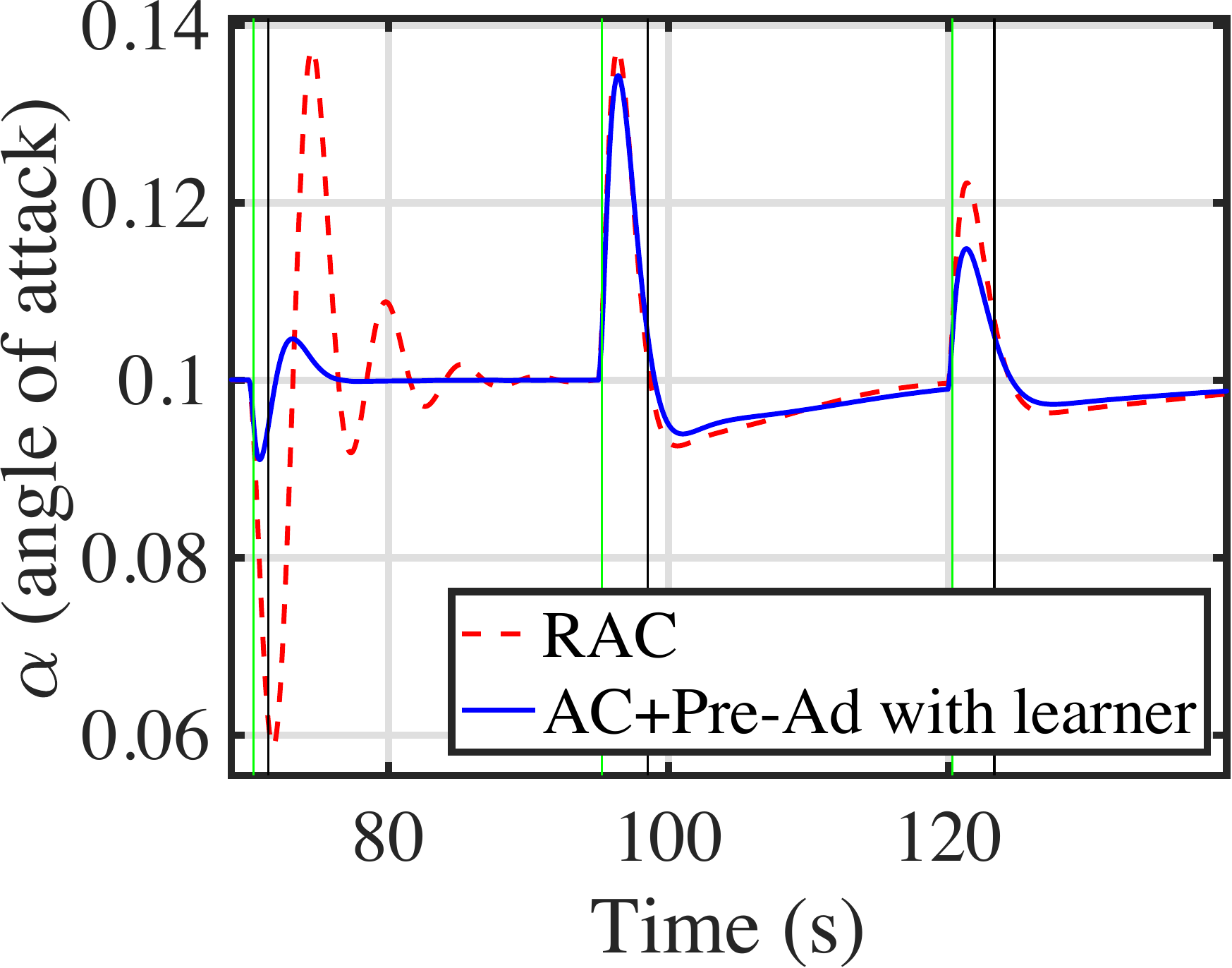} \\
\end{tabular}
\caption{Response of $\alpha$ (angle of attack) for the scenario in Eq. \eqref{eq:theta-scenario-2}. Left: response for $t \leq 140$, right: reponse for $70 \leq t \leq 140$. Green line: $E_u = 1$, black line: $E_d = 1$. RAC: regular adaptive control, AC+Pre-Ad with learner: adaptive control with the preadaptation mechanism and the preadaptation learner.}
\label{fig:scenario-response-2}
\end{figure}

Figure \ref{fig:scenario-response-2} gives the response for this scenario. As anticipated the response after the first drop, which happens at $t = 45$, is not better then the regular adpative control. We find that, at the next instant when the magnitude of the parameter drops, i.e. at $t = 70$, the response is much improved compared to the regular adaptive control, suggesting that the learning algorithm has been effective in updating the preadaptation function to respond to reductions in the parameter value. We also observe that, at the next instants, i.e. at $t = 90$ and $t = 120$, when the magnitude of the parameter increases again, the response after the disturbances continue to be better than the regular adaptive control suggesting that the preadaptation function has retained the memory of how to respond to increases in the parameter value.

%The algorithm has several hyperparameters which are the gain $\gamma_{pa}$ of the pre-adaption learner, the number of hidden layer neurons of the preadaptation function, and the thresholds of the attention function. These hyperparameters have to be tuned for the preadaptation mehanism to be effective. In the final manuscript we will discuss the selection of hyperparameter in more detail.

\begin{figure}
\centering
\begin{tabular}{ll}
\includegraphics[scale = 0.35]{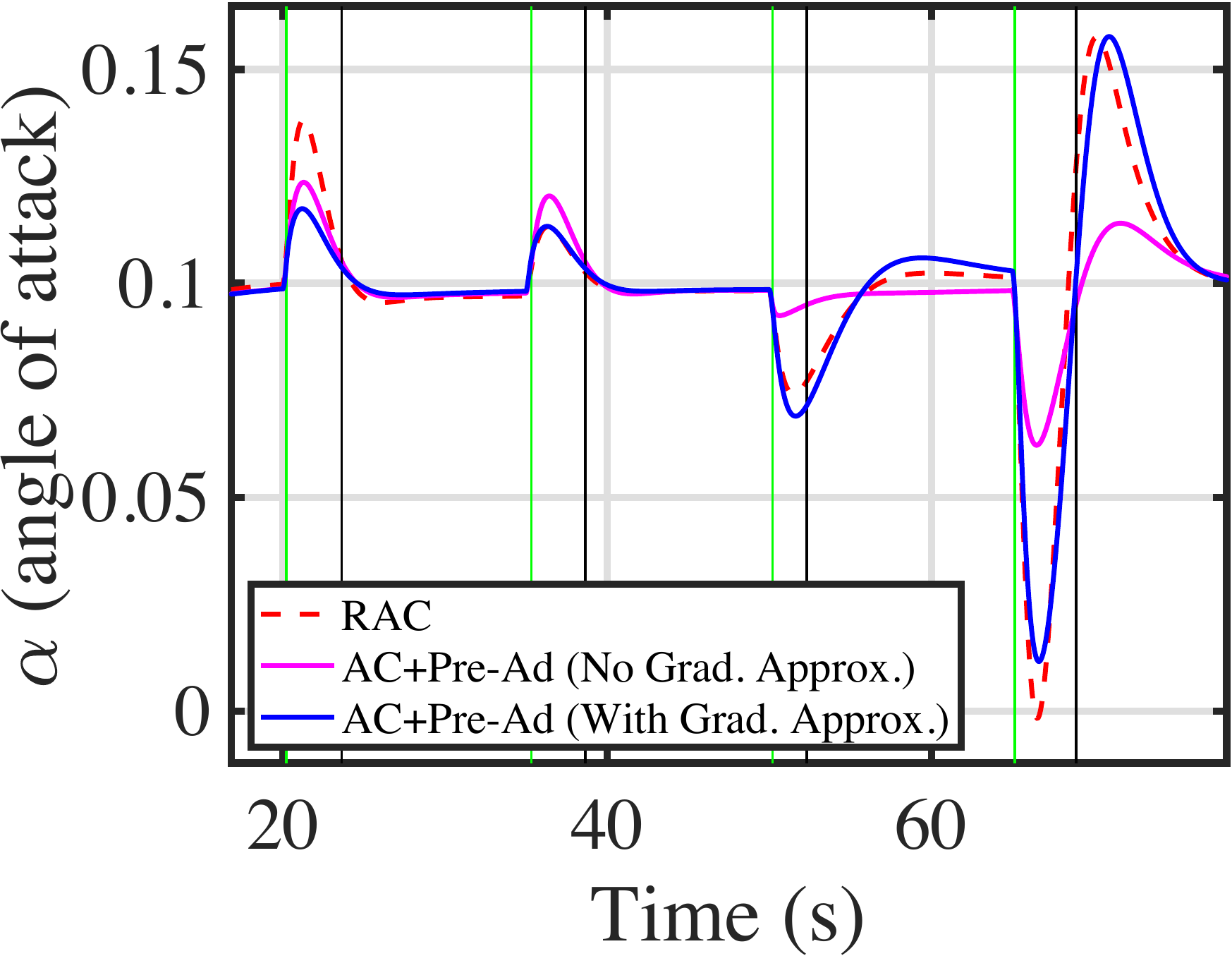} & \includegraphics[scale = 0.35]{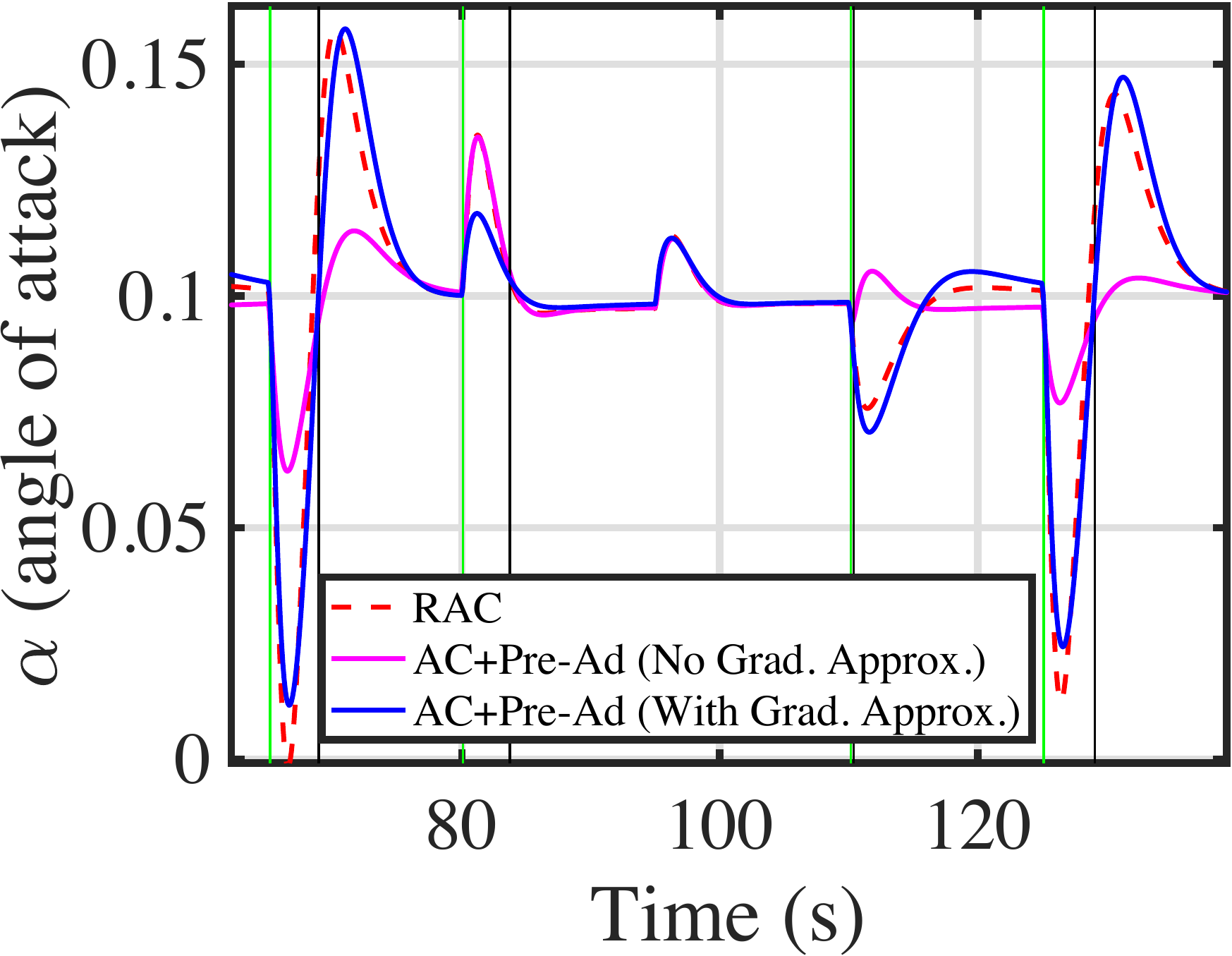} \\
\end{tabular}
\caption{Response of $\alpha$ (angle of attack) for the scenario in Eq. \eqref{eq:theta-scenario-3}. Left: response for $t \leq 80$, right: reponse for $65 \leq t \leq 140$. Green line: $E_u = 1$, black line: $E_d = 1$ for the preadaption learner without gradient approximation. RAC: regular adaptive control, AC+Pre-Ad: adaptive control with the preadaptation mechanism and the preadaptation learner.}
\label{fig:scenario-response-3}
\end{figure}

{\it Gradient Approximation}: Here we provide comparison between the preadaptation learner whose gradient is approximated as discussed in Section \ref{sec:cogpread} and the preadaptation learner without any approximation in its gradient based update. The scenario we consider to illustrate is a more complex scenario and is given by
\begin{align}
& \theta = 0.1\mathbf{1}, t \leq 5,  \ \ \theta = 1 \mathbf{1}, 5 < t \leq 20, \ \ \theta = 5\mathbf{1}, 20 < t \leq 35,  \ \ \theta = 10\mathbf{1}, 35 < t \leq 50, \nonumber \\
& \theta = 5\mathbf{1}, 50 < t \leq 65, \ \theta = 1\mathbf{1}, 65 < t \leq 80, \ \theta = 5\mathbf{1}, 80 < t \leq 95, \nonumber \\
& \theta = 10\mathbf{1}, 95 < t \leq 110, \ \theta = 5\mathbf{1}, 110 < t \leq 125, \ \theta = 1\mathbf{1}, 125 < t \leq 140.
\label{eq:theta-scenario-3} 
\end{align}
The difference in this scenario is that compared to scenario 2 the magnitude of the jumps are larger. Figure \ref{fig:scenario-response-3} provides a comparison of the response for the preadaptation mechanisms with and without gradient approximation in the learner. We find that the response for the preadaption learner with gradient approximation is similar to the pattern we had observed for the previous two scenarios, which is as expected. 

We find that the response for the preadaptation learner without any approximation is worser in some cases and better in other cases when compared to the preadaptation learner with the approximation in the gradient. We found this to be the case for the two scenarios we had considered earlier as well. This suggests that we cannot draw a clear conclusion whether the approximation of the gradient in the preadaptation learner affects the overall performance. We emphasize that further understanding on how the approximation affects the preadaptation learner and the response is required and this is a subject of future work. 

Overall, what we find is that both the preadaptation mechanisms result in improved performance in comparison to the regular adaptive controller. Another crucial observation is that the improved response did not result in high frequency oscillations, which are typically observed when the learning rates in regular adaptive control are increased.

{\it Hyperparameters}: The key hyperparameters are the learning rate $\gamma_{pa}$ and the thresholds $c_e$ and $c_{ed}$. The learning rate $\gamma_{pa}$ cannot be high or low. A higher learning rate is inappropriate because it will result in a very inaccurate preadaptation mechanism. At the same time a lower learning rate may not update the preadapation mechanism at all. So setting an appropriate value for the learning rate is essential. For the system we have presented, we found $\gamma_{pa} = \gamma =10$ to be a choice that was able to learn a preadaptation function that was effective. The thresholds are also critical because they determine the point at which the parameter to be adapted is reset by the preadaptation mechanism and the end point till which the adaptation phase that is used to update the preadaptation mechanism is recorded. The treshold values cannot be high because then the attention mechanism can miss detecting the onset of an adaptation phase. On the other hand, lower values are also not desirable because then smaller random fluctuations, which might arise from noise in the measurements, might get wrongly detected as the onset of disturbance. 

\section{Conclusion}

In this paper, we proposed a novel control architecture and algorithm for incorporating preadaption functions. We proposed a preadaptation mechanism that can augment any adaptive control scheme for a general linear system with linear parametric uncertainty. We showed that the preadaptation mechanism is effective in improving the adaptation across a wide range of scenarios and that it can reduce the peak of the response by as much as $50\%$ in some cases. We also proposed a preadaptation learner that learns the preadaptation function with experience, thus removing the complexity of designing and fine tuning the preadaptation function specific to the system to be controlled.

\section{Acknowledgement}

We thank the National Science Foundation (NSF) for supporting us under the NSF Grant Number ECCS-1839429.

\bibliography{Ref}

\end{document}